\documentclass[%
 reprint,
 amsmath,amssymb,
 aps,
]{revtex4-1}

\usepackage{graphicx}
\usepackage{dcolumn}
\usepackage{bm}

\begin{document}

\preprint{APS/123-QED}

\title{Lyth Bound, eternal inflation and future cosmological missions}

\author{Alessandro Di Marco}
\email{alessandro.di.marco@roma2.infn.it}
\affiliation{%
University of Rome - Tor Vergata, Via della Ricerca Scientifica 1\\
INFN Sezione di Roma “Tor Vergata”, via della Ricerca Scientifica 1, 00133 Roma, Italy}%


\begin{abstract}

In this paper we provide a new expression for the variation of the inflaton field $\Delta\phi$ during the horizon crossing epoch in
the context of single field slow roll inflationary models.
Such an expression represents a generalization of the well-know Lyth bound.
We also explore the consequences of a detection of permille order of the tensor-to-scalar ratio amplitude, $r$, as well as an improvement
on the estimation of the scalar spectral index, $n_s$ and its running $\alpha_s$, by the upcoming cosmic macrowave background (CMB) polarization experiments
that will provide plausible constraints on the quantity $\Delta\phi$ during the horizon exit moment.
In addition we discuss the relation between the local variation of the field and the possibilities of an eternal inflation.
The results of the analysis are completely model independent.

\end{abstract}

\pacs{Valid PACS appear here}
\maketitle


\section{Introduction}

The inflationary stage is the widely accepted mechanism to explain the physics of the early universe \cite{1}.
The simplest and most promising version involves a single field slow-roll inflation where an ordinary scalar field $\phi$ (neutral, homogeneous, minimally coupled to gravity and canonically normalized) explores an appropriate effective potential $V(\phi)$
to realize a quasi de-Sitter evolution of the spacetime, if the slow roll condition $\dot{\phi}^2\ll V(\phi)$ holds long enough.
This scalar field is often called inflaton field and the function $V(\phi)$ is the inflationary potential.
Although inflation explains several features of our universe (absence of monopoles, flatness, homogeneity and isotropy, adiabatic
and scale invariant fluctuations) we need to address and resolve compelling open issues.
One of these is the nature of the scalar field $\phi$ and the size of the distance covered by $\phi$ during inflation, $\Delta\phi$.
Several authors used to say that a trans or super-planckian excursion of the inflaton field, $\Delta\phi>M_p$ could be very problematic
especially from the standpoint of the low energy effective field theory (EFT) with a given Planck cut-off \cite{2,3,4}.
However, in the last years it has been shown that the question is not general: the well-known $\alpha$-attractor models of inflation
do not suffer from this problem.
In any case, Lyth derived a bound (see \cite{2}) for the value assumed by $\Delta\phi$ in terms of the amplitude $r$ of the primordial gravitational waves produced
during inflation (in the context of slow-roll inflationary models).
This quantity is known as Lyth Bound and represents an estimation of $\Delta\phi$ at horizon crossing: $\Delta\phi_{H.c.}$.
In the last years, different studies concerning the variation of the field and the Lyth bound itself have been done \cite{5,6,7}.
In future, foreseen CMB polarization missions (\cite{8,9,10,11,12}) and gravitational waves experiments (\cite{13}) will constrain cosmological observables better than the current available data, \cite{14,15}.
In particular, they will be able to explore values of $r$ of the order of $10^{-3}$ and to reduce the uncertainty on $n_s$ and $\alpha_s$ better than a factor 3.
Therefore, in this paper we provide a new slow-roll expansion of the standard Lyth bound to take into account our (current and foreseen) knowledge about the scalar
spectral index, $n_s$, and the running of the scalar spectral index, $\alpha_s$.
Then, we present plausible constraints on $\Delta\phi$ at horizon crossing with respect to future cosmological results.
The estimation of the running of the scalar perturbations, $\alpha_s$, could be important to evaluate the possibility of an eternal inflation (see \cite{16,17,18,19}), as outlined by W.Kinney and K.Freese in \cite{20}.
Therefore, we will discuss the implication of an eternal inflation on the local variation of the field.
The paper is organized in the following way.
In Sec. II we summarize the problem of the variation of the field and we introduce the standard definition of Lyth Bound and finally, we use the inflationary flow equations to generalize the Lyth formula.
In Sec. III we discuss what we can expect on $\Delta\phi_{H.c.}$ in view of future polarization missions and we explore the relation between eternal inflation and $\Delta\phi_{H.c.}$. 
Finally, Sec. IV is dedicated to our findings.
In this manuscript we use the particle natural units, $c=\hbar=1$, $m_p$ indicates the Planck mass and $M_p$ is the reduced Planck mass,
with $M^2_p=m_p^2/8\pi$.

\section{Slow roll inflation and the variation of the scalar field} 

\subsection{Effective field theory, super-planckian excursions and Lyth Bound}

The dynamics of a slow roll inflationary universe can be described within a Hamilton-Jacobi picture
by the following system of coupled equations:
\begin{eqnarray}\label{eqn: equazione HJ}
V(\phi)=3M^2_p H^2(\phi) - 2 M^2_p  H'(\phi),
\end{eqnarray}
\begin{eqnarray}\label{eqn: fieldeq}
\dot{\phi}=-2 M^2_p H'(\phi)                
\end{eqnarray}
where we can assume $\dot{\phi}<0$ and $H'(\phi)>0$.
As remarked in Sec. I, we have a quasiexponential evolution of the spacetime.
At the same time, the scalar field gets a slow evolution (over the first stage of inflation) because $\dot{\phi}^2\ll V(\phi)$.
At first order, the resulting dynamics is well described by the first two Hubble slow roll parameters:
\begin{eqnarray}\label{eqn: hsrp}
\epsilon(\phi)=2M^2_p\left( \frac{H'(\phi)}{H(\phi)}  \right)^2, \quad   \eta(\phi)=2M^2_p\left( \frac{H''(\phi)}{H(\phi)}  \right)
\end{eqnarray}
where $\sqrt{\epsilon}>0$ if $\dot{\phi}<0$.
The nature of the scalar field $\phi$ is an open issue: it could be a fundamental particle field (for instance, an Higgs field \cite{20}) or an
auxiliary field emerging from a modification of gravity (for example, Starobinisky $R^2$ model or its extensions \cite{21,22}).
In any case it is common believed that the inflaton could be an effective light scalar degree of freedom in a low energy-limit description of some more fundamental quantum gravity cosmological theory.
This piece of evidence led people to make the following observations.
We can think inflation as emerging from a low energy effective field theory (EFT) with some Planck cut-off $\Lambda$.
Therefore, the corresponding effective (slow roll) inflationary lagrangian is obtained integrating out the Planck-scale (quantum gravity) degrees of freedom (see \cite{23} for a review).
As a result, the
slow roll effective potential $V(\phi)$ receives an infinite number of corrections, given by Planck suppressed operators $\mathcal{O}^{(n)}$ \cite{2,3,4}

\begin{eqnarray}\label{eqn: eft}
\mathcal{L(\phi,\partial_{\mu}\phi})&=&-\frac{1}{2}\partial_{\mu}\phi\partial^{\mu}\phi - V(\phi)\left[ 1 + \sum_{n=1}^{\infty} \frac{\mathcal{O}^{(n)}}{M_p^{n-4}} \right]\\ \nonumber
&=&-\frac{1}{2}\partial_{\mu}\phi\partial^{\mu}\phi - V(\phi)\left[ 1 + \sum_{n=1}^{\infty} \lambda_{n}\left(\frac{\phi}{M_p}\right)^n \right].
\end{eqnarray} 

In particualar, the operators $\mathcal{O}^{(n)}$ result in power of $\phi$.
The $\lambda_{n}$ are the Wilson coefficients, typically of order $1$.
This means we have an infinite number of corrections of the same order.
Now, as stressed by different authors (see \cite{2,3}), if the scalar field assumes subplanckian values, $\phi\ll M_p$, i.e is small, 
the series is asymptotically convergent.
This suggest we can describe inflation by an EFT whether $\Delta\phi\ll M_p$ over the entire inflationary history.
On the other side, if the field takes on trans or super-Planckian values, $\phi>M_p$, the series is divergent.
Then, under this condition, an EFT-description of inflation it is hard to realize because
higher-order operator terms become important and the balance between the shape and the height of the potential is not guaranteed (see again \cite{2,3}).
For example, operator terms of the form:
\begin{eqnarray}
\Delta V_{2}(\phi)\sim V(\phi)\frac{\phi^2}{M_p^2}
\end{eqnarray}
contributes to the slow roll parameter $\eta$ with order $1$ not providing a sufficient amount of inflation (this is the so called $\eta$-problem, \cite{23}).
At this point, as summarized in \cite{23}, we can make a sense to the EFT slow roll models only if the Wilson coefficients are small, $\lambda_n\ll 1$.
This is a distinct possibility if we introduce a symmetry for the effective lagrangian with the prescription that 
such a symmetry is also recognized in the UV completion of our EFT (e.g. in superstring theory, \cite{24,25}) to ensure that it is also realized even at the fundamental Planck-scale \cite{26}.
A typical example is the shift symmetry.
Indeed shift symmetry is not broken by perturbative corrections of quantum gravity.
In principle, it could be broken by nonperturbative effects although they are so small to result negligible \cite{27}.
Actually, other authors like Linde (see \cite{28} for a review), state that the problem of $\Delta\phi$-size and therefore the values assumed by the scalar field
is a relative question.
In fact, one should just make sure that the energy density is sub-Planckian, $V(\phi) < M^4_p$ in order to avoid quantum gravity effects (while the field can still assumes larger values in Planck units).
However, the current state of the art is a little bit different.
For instance, in the last years A.Linde, R.Kallosh and others developed a new important class of inflationary models called $\alpha$-attractors models
which the most advanced version is realized in the supergravity framework \cite{29}.
These models are characterized by an asymptotic flatness region for large value of $\phi$ and they interpolate a broad range of predictions in the $(n_s,r)$-plane. 
Furthermore, they do not suffer from the problem discussed before because the asymptotic flatness of the inflationary potential is preserved by
dangerous quantum corrections as shown in \cite{30}.
However, the excursion of the scalar field $\Delta\phi$ over the inflationary phase shows a lower limit thanks to Lyth \cite{2}.
This limit can be derived using different approach. One way is to consider the Eq.($\ref{eqn: fieldeq}$)
and rewrite the derivative with respect to time in a derivative with respect to the number of $e$-foldings $N$.
As a result, we get:
\begin{eqnarray}\label{eqn: basiceq}
\frac{d\phi}{dN}=\sqrt{2}M_p\sqrt{\epsilon(\phi)}.
\end{eqnarray}
This equation is particularly important because allows to reword a $\phi$-type derivative in a $N$-type derivative.
Note that this quantity is modulated by the first slow roll parameter $\epsilon$.
The slow roll condition for the inflaton dynamics implies $r\sim 16\epsilon$ and both of these parameters are slowly varying.
Then, the variation of the field along the horizon crossing period can be written as:
\begin{eqnarray}
\frac{\Delta\phi_{h.c.}}{M_p}=\sqrt{\frac{r}{8}}|\Delta N| \quad \mbox{ or}\quad \frac{\Delta\phi_{h.c}}{m_p}=\sqrt{\frac{r}{64\pi}}|\Delta N|.
\end{eqnarray}
Nevertheless, the observable scales from which we deduce the value of or the bounds for the inflationary parameters like $n_s,r,\alpha_s,...$,
correspond to multipoles $2< l <100$.
These scales leave the Hubble horizon $R_H=1/H$ along a period of $\Delta N\sim 4$ $e$-foldings.
Therefore, we can state that the total excursion of the scalar field is larger than:
\begin{eqnarray}
\frac{\Delta\phi}{m_p}\geq\sqrt{\frac{r}{4\pi}}.
\end{eqnarray}
This result is model dependent in the sense that, each model of inflation is characterized by a minimal excursion modulated by the amplitude $r$.
Let us assume that the number of $e$-foldings between the horizon crossing epoch and the end of inflation epoch is $N_*\sim 60$.
Then, for a quadratic potential $V(\phi)\sim m^2\phi^2$ we have:
\begin{eqnarray}
\frac{\Delta\phi_{h.c.}}{m_p}\sim 0.1
\end{eqnarray}
while for a quartic potential $V(\phi)\sim \lambda\phi^4$ the corresponding variation is
\begin{eqnarray}
\frac{\Delta\phi_{h.c.}}{m_p}\sim 0.14.
\end{eqnarray}
Note that, we have to keep in mind that different potentials can produce the same $r$ and so the same $\Delta\phi$.
On the other hand, a precise detection of the tensor-to-scalar ratio plays an important role to determine
the minimal displacement of the field beyond the specific model.
However, the Lyth Bound is a first order result.
In the future, different cosmological missions aim to improve the current
knowledge about the inflationary parameters.
In this respect, we want to generalize the Lyth Bound in order to include the $n_s$ and $\alpha_s$ contributions and exploring so
possible constraints on $\Delta\phi_{H.c.}$ at next orders.

\subsection{Generalizing Lyth Bound}

In this section we aim to derive a higher order expression of the Lyth Bound.
To reach this goal, the slow roll parameters and the related inflationary flow equations are the fundamental tools.
The Hubble slow roll parameters are defined by the following hierarchy of variables \cite{31}:
\begin{eqnarray}
\epsilon(\phi)&=&2M^2_p\left(\frac{H'}{H} \right)^2 \\
\beta^{(n)}(\phi)&=&(2M^2_p)^n \left[ \frac{H'^{n-1} H^{(n+1)}}{H^n}  \right]
\end{eqnarray}
where $\eta=\beta^{(1)}$, $\xi^2=\beta^{(2)}$, $\sigma^3=\beta^{(3)}$ ... with $\sqrt{\epsilon}>0$ for $\dot{\phi}<0$, as remarked in Sec. II.
The slow roll parameters are not constant in general but they change as the scalar field evolves.
However, as explained by Easther, Kinney and Powell in \cite{3}, the evolution of the slow roll parameters could be very slow (for example during the horizon crossing epoch) in some phases of inflation compared to other moments (the last $e$-folds of inflationary expansion).
Mathematically, the evolution of the $\beta_i(\phi)$ are well-described by the system of inflationary flow equations, in which, for sake of convenience,
the slow roll parameters are suitably expressed as functions of the number of $e$-foldings.
In particular, we can write
\begin{eqnarray}
\frac{d\epsilon}{dN}&=&2\epsilon(\eta -\epsilon) \\
\frac{d\eta}{dN}&=&\xi^2 -\epsilon\eta\\
\frac{d\beta^{(n)}}{dN}&=&\left[(n-1)\eta - n\epsilon  \right]\beta^{(n)} + \beta^{(n+1)}.
\end{eqnarray}
Greater details can be found in references \cite{32,33}.
At this point, we can follow the method used in \cite{3} and write down a Taylor expansion of the scalar field (as a function of the number of $e$-foldings)
around the horizon crossing of the observable scales
\begin{eqnarray}
\phi(N)=\phi(N_*) + \phi'(N_*)\Delta N + \frac{1}{2!}\phi''(N_*)\Delta N^2 + ...
\end{eqnarray}
where $'$ indicates a derivative with respect to $N$ and $\Delta N= N-N_*$
Assuming again that $\phi$ is decreasing along the inflationary evolution and that $N_*>N$, we can rewrite the Taylor expansion as follows:
\begin{eqnarray}
\Delta\phi= + \phi'(N_*)|\Delta N| - \frac{1}{2!}\phi''(N_*)|\Delta N|^2 + ...
\end{eqnarray}
where $\Delta\phi=\phi_*-\phi$ and $|\Delta N|=-(\Delta N)$.
Generalizing, we have
\begin{eqnarray}\label{eqn: exp}
\Delta\phi=\sum_{n=1}^{\infty} c_n |\Delta N|^n, \quad c_n=(-1)^{n+1} \phi^{(n)}(N_*).
\end{eqnarray}
The first order derivative of the expansion is the previous basic equation, Eq.($\ref{eqn: basiceq}$)
\begin{eqnarray}\label{eqn: basiceq}
\frac{d\phi}{dN}=\sqrt{2}M_p\sqrt{\epsilon(\phi)}.
\end{eqnarray}
Using the flow equations we can derive the higher order derivatives of the field in terms of the slow roll parameters.
For example, next orders result in
\begin{eqnarray}\label{eqn: second}
\frac{d^2\phi}{dN^2}=\sqrt{2}M_p\sqrt{\epsilon}\left(\eta-\epsilon\right),
\end{eqnarray}
\begin{eqnarray}\label{eqn: third}
\frac{d\phi^3}{dN^3}=\sqrt{2}M_p\sqrt{\epsilon}\left(3\epsilon^2 + \eta^2  -5\epsilon\eta + \xi^2 \right)
\end{eqnarray}
and
\begin{eqnarray}\label{eqn: fourth}
\frac{d\phi^4}{dN^4}=\sqrt{2}M_p\sqrt{\epsilon}\left[f_1(\epsilon,\eta,\xi^2) + f_2(\epsilon,\eta,\xi^2,\sigma^3) \right]              
\end{eqnarray}
where
\begin{eqnarray}
f_1(\epsilon,\eta,\xi^2)=(\eta-\epsilon)(15\epsilon^2+\eta^2-15\epsilon\eta+\xi^2),
\end{eqnarray}
\begin{eqnarray}
f_2(\epsilon,\eta,\xi^2,\sigma^3)=(\xi^2-\epsilon\eta)(2\eta-5\epsilon) \nonumber \\ + \xi^2(\eta-2\epsilon) + \sigma^3&.
\end{eqnarray}
Here $\sigma^3$ is the fourth slow roll parameter.
At this point, we can calculate the derivatives at horizon crossing moment $t_*$ (or $N_*$, let us say).
In doing so, we have to reword the slow roll parameters in terms of the inflationary observables.
At first order, we have
\begin{eqnarray}
\epsilon\sim \frac{r}{16}, \quad \eta\sim\frac{1}{2}(n_s-1) + \frac{r}{8}
\end{eqnarray}
and
\begin{eqnarray}
\xi^2 \sim 5\epsilon\eta - 4\epsilon^2 - \frac{1}{2}\alpha_s.
\end{eqnarray}
Substituting these information in the expansion Eq.($\ref{eqn: exp}$) with $|\Delta N|\sim 4$ $e$-folds, we have (at the third order, for instance)
\begin{eqnarray}
\frac{\Delta\phi_{h.c}}{m_p}\sim\sqrt{\frac{r}{4\pi}}\Big\{ \Pi_0-\Pi_1 - \Pi_2 + ... \Big\}
\end{eqnarray}
where:
\begin{eqnarray}
\Pi_0&=&1\\
\Pi_1&=&(n_s-1)+\frac{r}{8}\\
\Pi_2&=&\frac{8}{3}\left(\frac{r}{16} \right)^2-\frac{8}{3}\left[ \frac{1}{2}(n_s-1) + \frac{r}{8} \right]^2 + \frac{4}{3}\alpha_s
\end{eqnarray}
where $\Pi_1=\Pi_1(r,n_s)$, $\Pi_2=\Pi_2(r,n_s,\alpha_s)$.
The first term of the expansion is the standard Lyth Bound.
The second order term, linear in $n_s$, is the result found by R.Easther et all in \cite{3}.
Finally, the last term is the new corrections that take into account the contribution of the running of the scalar spectral index and the quadratic order in $r$
and $n_s$.
Note that, the contribution of $\alpha_s$ in our formula appears with a negative sign
\begin{equation}\label{eqn: alphadep}
\frac{\Delta\phi_{h.c}}{m_p}\sim -\frac{4}{3}\alpha_s
\end{equation}
Therefore a negative value for $\alpha_s$ pushes the local variation $\Delta\phi$ to larger values.
On the contrary, if $\alpha_s$ results to be positive, the excursion receives a negative contribution.
In the next section, we use our third order result to explore the constraints we may have on the variation $\Delta\phi_{h.c}$ 
in terms of the expected results of next generation experiments and we discuss the implication of an eternal inflation on the variation of the field.

\section{What we can expect by upcoming cosmological missions?}

\subsection{Monte Carlo simulation}

\begin{figure}[htbp]
\centering
\includegraphics[width=8.5cm, height=6cm]{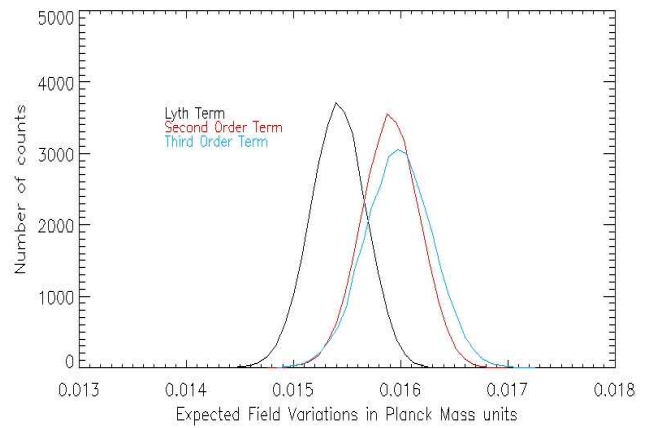}
\caption{Distribution functions for the minimum excursion of the scalar field during inflation, $\Delta\phi_{h.c}$, for the three different slow roll expansions.
Here, we have supposed a tensor-to-scalar ratio amplitude of the order of $\mu_{r}\sim 3\times 10^{-3}$ with $n_s=0.9680$, $\alpha_s=-0.0033$}
\label{fig: 1}
\end{figure}

\begin{figure}[htbp]
\centering
\includegraphics[width=8.5cm, height=6cm]{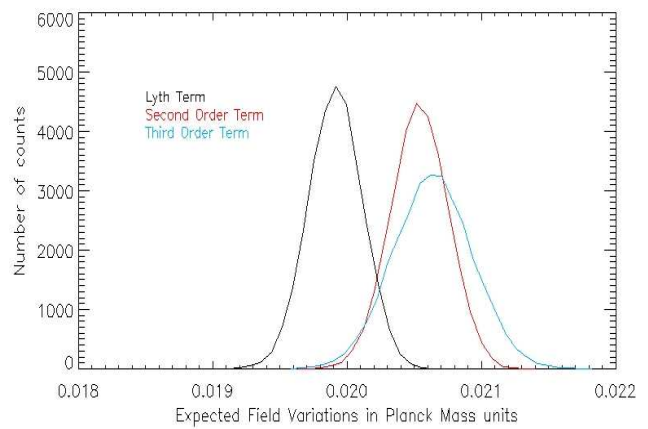}
\caption{Distribution functions for the variable $\Delta\phi_{h.c}$ for the three different slow roll expansions.
In this case, we have supposed a detection of primordial GW of the order of $\mu_{r}\sim 5\times 10^{-3}$ with $n_s=0.9680$, $\alpha_s=-0.0033$}
\label{fig: 2}
\end{figure}

\begin{figure}[htbp]
\centering
\includegraphics[width=8.5cm, height=6cm]{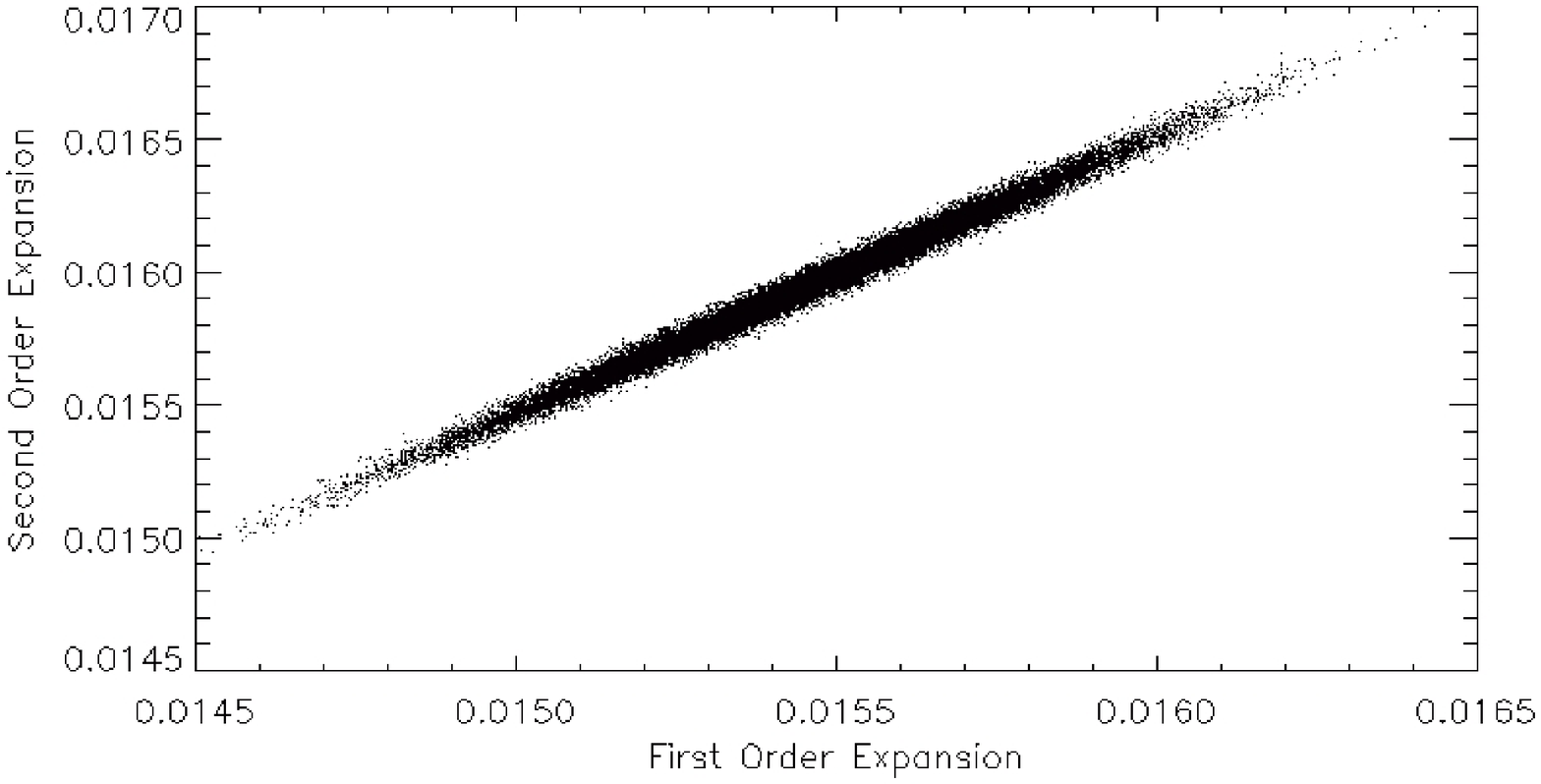}
\caption{Dispersion plot for the ratio $\Delta\phi(\Pi_1)/\Delta\phi(\Pi_0)$ in the case of a detection of $r\sim 3\times 10^{-3}$ with $n_s=0.9680$, $\alpha_s=-0.0033$.}
\label{fig: 3}
\end{figure}

\begin{figure}[htbp]
\centering
\includegraphics[width=8.5cm, height=6cm]{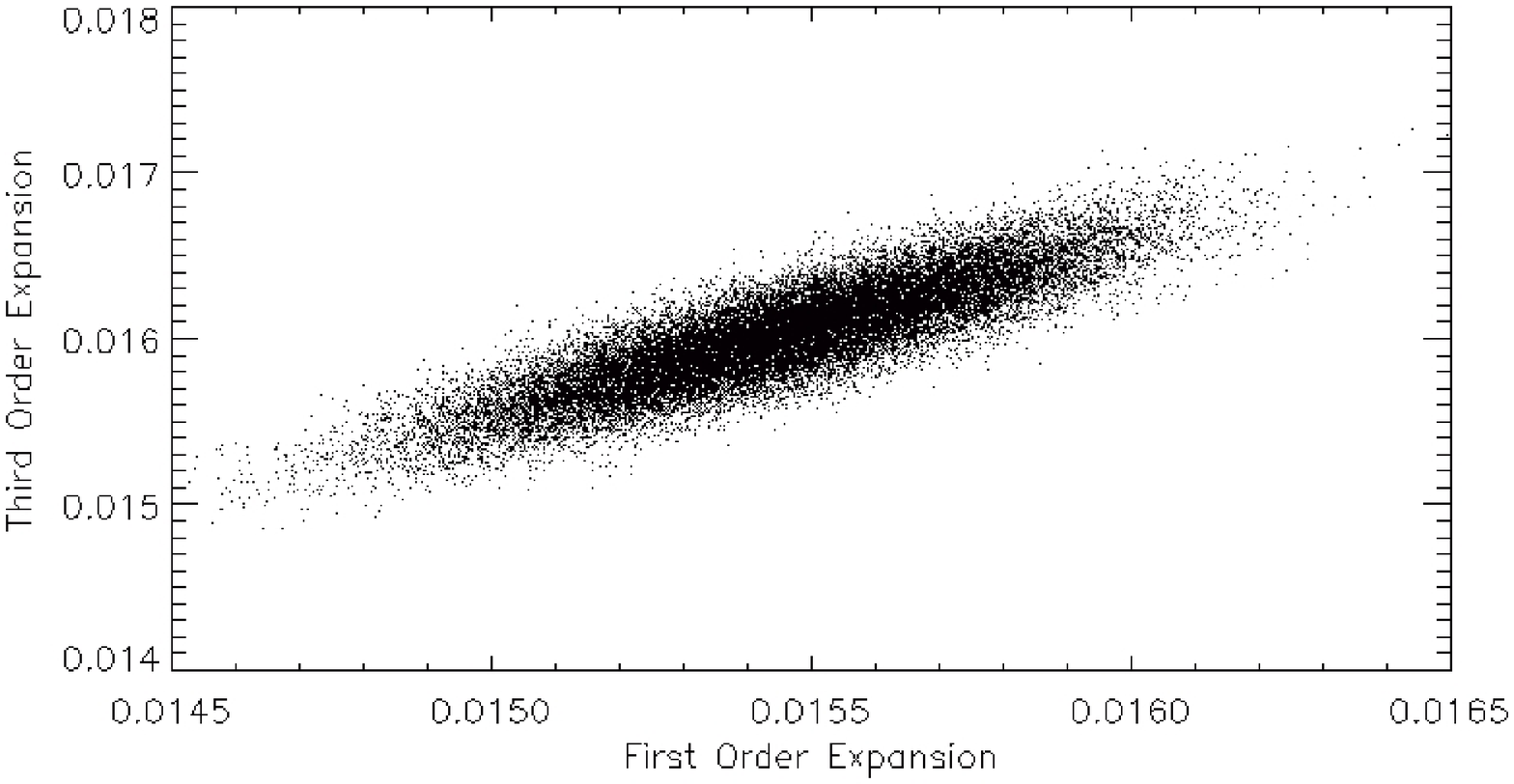}
\caption{Dispersion plot for the ratio $\Delta\phi(\Pi_2)/\Delta\phi(\Pi_0)$ in the case of a detection of $r\sim 3\times 10^{-3}$ with $n_s=0.9680$, $\alpha_s=-0.0033$.}
\label{fig: 4}
\end{figure}

The current state-of-the art for the values of the main inflationary parameters is the following (for detailed papers see \cite{11,12}:
\begin{eqnarray}
n_s=0.9680\pm 0.006, \quad r<0.07 \mbox{ at 95\% C.L.}
\end{eqnarray}
and 
\begin{eqnarray}
\alpha_s=\frac{dn_s}{d\ln k}=-0.0033\pm 0.0074
\end{eqnarray}
However, the above limits could rapidly improve in the close future due to the next generation of cosmological experiments, like
polarization missions of the cosmic macrowave background (CMB) or gravitational waves (GW) detection missions.
These new missions aim to improve the knowledge on $n_s$ and $\alpha_s$ and to probe scales of the order of $10^{-3}$ for the tensor-to-scalar ratio amplitude, $r$
(or better in the best scenario, \cite{8,9,10,11,12,13}).
In this regard, we investigate the variation of the field over the first $4$ $e$-folds (more or less)
of inflationary expansion.
To do this, we consider a multivariate Gaussian $\mathcal{G}(r,n_s,\alpha_s)$ distribution for the parameters $(r,n_s,\alpha_s)$ with
\begin{enumerate}
\item $\mu_{r}=0.003,0.004,0.005$ 
\item $\mu_{n_s}=0.9620,0.9650,0.9680$ 
\item $\mu_{\alpha_s}=-0.0005,-0.0015,-0.0025$  
\end{enumerate}
with uncertainties $\sigma_r=0.0001$, $\sigma_{n_s}=0.002$, $\sigma_{\alpha_s}=0.0023$, respectively.
Furthermore, we set to zero (for simplicity) all the correlation coefficients: $\rho_{rn_s}=\rho_{r\alpha_s}=\rho_{n_s\alpha_s}=0$.
In Table I, we report the simulation results for the size of $\Delta\phi_{h.c}$ for the three different assumed values of the tensor-to-scalar ratio.
In this case, we report the results for all of the three expressions $\Delta\phi_{h.c}(\Pi_i)$.
On the other side, in Table II and in Table III we skip the term $\Delta\phi_{h.c}(\Pi_0)$ and $\Delta\phi_{h.c}(\Pi_1)$, respectively.
This is due to the structure of the expressions.
Furthermore, Fig.($\ref{fig: 1}$) and in Fig.($\ref{fig: 2}$) show the resulting distribution functions for the minimum value of $\Delta\phi$ supposing a detection of primordial tensor modes given by a forthcoming CMB baloon polarization mission or GW mission while Fig.($\ref{fig: 3}$) and Fig.($\ref{fig: 4}$) report the associated dispersion relations
$\Delta\phi_{h.c}(\Pi_1)/\Delta\phi_{h.c}(\Pi_0)$ and $\Delta\phi_{h.c}(\Pi_2)/\Delta\phi_{h.c}(\Pi_0)$, respectively.
In both cases, we assume the current data for the sampling of $n_s$ and $\alpha_s$.
Now, these Monte-Carlo simulations show a couple of properties.
First of all, we can deduce from Table I the inclusion of the scalar spectral index $n_s$ and the running $\alpha_s$ produces a larger $\Delta\phi_{h.c}$.
Second, the $1$-$\sigma_{\Delta\phi_{h.c}}$-value turns out to be larger as the slow roll reconstruction become deeper.
Somehow, this is an expected result and it is originated by the introduction of more uncertainty due to the new parameters. 
On the other hand, Table II tells us that the variation of the field become smaller as $n_s$ increases.
In fact, the scalar spectral index, always appears in the $\Delta\phi_{h.c}$ expression like $1-n_s$.
Therefore, when $n_s$ gets larger values, the previous term approaches to zero so it is a subdominant contribution.
The last table (Table III) shows that a bigger running index (in modulus) implies a larger excursion because of Eq.($\ref{eqn: alphadep}$).

\subsection{Lyth Bound and eternal inflation}

\begin{figure}[htbp]
\centering
\includegraphics[width=8.5cm, height=6cm]{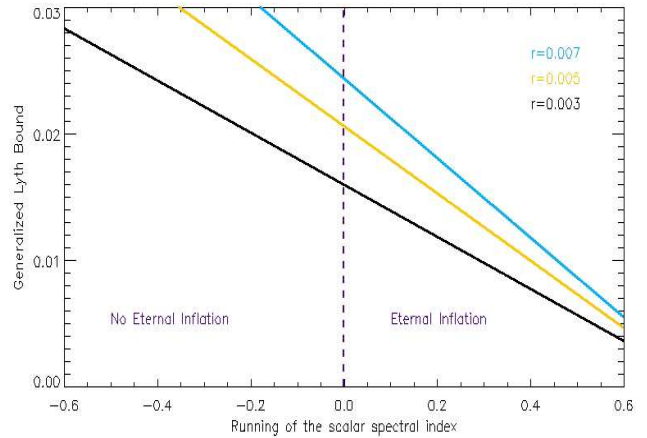}
\caption{Third order Lyth Bound as a function of the running $\alpha_s$. In this case we consider $n_s\sim0.9680$ with a threshold value $\alpha_s\sim-1.2842\times 10^{-5}$. Possible eternal inflation phases are associated with smaller variations of the field at horizon crossing of quantum fluctuations.}
\label{fig: deltaphi_eternal}
\end{figure}

The inflationary mechanism could lead to a very interesting and spectacular consequence: an infinitely self-reproducing state commonly called eternal inflation.
The eternal inflation process was firstly outlined by A.Linde, P.Steinhardt and A.Vilenkin (see \cite{16,17}) in the framework of new inflation.
Afterwards, A.Linde realized that such a mechanism could be also possible in the chaotic inflationary scenario \cite{18} (see also, \cite{19} for greater details).
In this second case, the eternal inflation stage is laid down when the inflaton field explores a region $\Sigma$ of the inflationary potential in which quantum fluctuations of the field $\delta\phi$ are larger then the classical variation ones, $\Delta\phi$.
This leads to a very interesting implications about the amplitude of the metric perturbations as pointed out by A.Linde in the second of references \cite{19} 
and summarized in \cite{34,35}: the power spectrum of the scalar metric perturbations must exceed the unity on those physical scales related to $\Sigma$
\begin{equation}\label{eqn: eternal condition}
P_s(k)=A_s\left(\frac{k}{k_*}\right)^{f(k,k_*)}>1 \mbox{ for all } k\in \mathcal{I}_{\Sigma}(k)
\end{equation}
where $\mathcal{I}_{\Sigma}(k)$ is the set of scales $k$ related to the portion $\Sigma$ of inflationary potential,
$k_*$ is the pivot scale for probing the cosmological parameters and the function $f(k,k_*)$ is given by
\begin{equation}\label{eqn: f}
f(k,k_*)=(n_s-1)+\alpha_s\ln\left(\frac{k}{k_*}\right)+\beta_s\ln^2\left(\frac{k}{k_*}\right)+...
\end{equation}
in which $\beta_s$ and the higher order terms represent the running-of-the-running, the running-of-the-running-of-the-running and so on.
Who is $I_{\Sigma}(k)$?
We should note that the current cosmological results show a very low $A_s$ ($A_s\simeq 2.2 \times 10^{-9}$, see \cite{14,15}) therefore
we can be fairly confident the observable and accessible region of $V(\phi)$ does not provide an eternal inflationary epoch. 
Then, we can expect an increase (at least) of the order of $10^9$ on $P_s(k)$ only at extremely large scales $k\ll k_*$, beyond our horizon in principle.
How can we get $P_s(k)\geq1$?
In general, we can get such a condition by proper choices of the function Eq.($\ref{eqn: f}$), i.e., by proper choice of the cosmological parameters.
Then, if we truncate the series at higher orders, the condition Eq.($\ref{eqn: eternal condition}$) is provided by a large number of choices, generally.
However, we may imagine that the power spectrum is modulated only by the scalar spectral index and by a non null and constant $\alpha_s$:
\begin{equation}
\alpha_s\neq 0, \quad \mbox{ and } \quad  \frac{d\alpha_s(k)}{d\ln k}=0
\end{equation} 
From this point of view, one can use the running as the only degree of freedom or ``temperature parameter" to study the transition to a blue spectrum $P(k)>1$.
We can call the threshold value of the running preventing the eternal inflation, $\alpha^*_s$ and find \cite{35}:
\begin{eqnarray}
\alpha^*_s=\frac{(1-n_s)^2}{4\ln A_s}
\end{eqnarray}
Therefore, we may argue that:
\begin{eqnarray}
\alpha_s&<&\alpha^*_s \mbox{ eternal inflation does not occur }\\
\alpha_s&>&\alpha^*_s \mbox{ eternal inflation occurs }
\end{eqnarray}
For instance,
\begin{eqnarray}
n_s\sim0.9620, \quad \alpha^*_s\sim-1.8109\times 10^{-5} \\ 
n_s\sim0.9650, \quad \alpha^*_s\sim-1.5363\times 10^{-5} \\ 
n_s\sim0.9680, \quad \alpha^*_s\sim-1.2842\times 10^{-5} 
\end{eqnarray}
In the previous section we derived an expression of the Lyth Bound in terms of $\alpha_s$.
Then, we can distinguish which are the minimum variations of the field related to an eternal inflation (EI) stage with respect to those are not:
\begin{eqnarray}
\Delta\phi_{h.c}(\alpha_s&<&\alpha^*_s) \mbox{ no EI-related variations }\\
\Delta\phi_{h.c}(\alpha_s&>&\alpha^*_s) \mbox{ EI-related variations }
\end{eqnarray}
In Fig.($\ref{fig: deltaphi_eternal}$) we report the third order expression of $\Delta\phi_{H.c.}$ as a function of $\alpha_s$ for different detections of primordial gravitational waves.
In addition, we outline the range of the possible EI-related $\Delta\phi_{H.c.}$ at horizon crossing of quantum modes.
In the next section we will discuss the reasonableness of these discussions.

\begin{table*}
\begin{ruledtabular}
\begin{tabular}{ccccc}
                                                           
 $\mu_{r}$        &          $\Delta\phi_{h.c}(\Pi_0)$  &  $\Delta\phi_{h.c}(\Pi_1)$  &  $\Delta\phi_{h.c}(\Pi_2)$    \\ \hline     
 $0.003$          &          $0.01545\pm0.00026$   &      $0.01598\pm0.00027$    &       $0.01605\pm0.00028$    \\      
 $0.004$          &          $0.01784\pm0.00022$   &      $0.01846\pm0.00024$    &       $0.01853\pm0.00024$     \\      
 $0.005$          &          $0.01995\pm0.00019$  &      $0.02063\pm0.00021$    &        $0.02072\pm0.00022$     \\      
\end{tabular} 
\end{ruledtabular}
\caption{Simulation results for the variation of the inflaton field at horizon crossing for the three different order of slow roll approximation.
Here, the scalar spectral index and the relative running are fixed to $n_s=0.9650$ and $\alpha_s=-0.0025$, respectively.
As one can see, the introduction of $n_s$ and $\alpha_s$ implies a larger excursion.}
\label{tab: table1}
\end{table*}

\begin{table}
\begin{ruledtabular}
\begin{tabular}{ccccc}
                                                           
 $\mu_{n_s}$       &           $\Delta\phi_{h.c}(\Pi_1)$       &    $\Delta\phi_{h.c}(\Pi_2)$     \\ \hline     
 $0.9620$          &           $0.02069\pm0.00021$       &    $0.02078\pm0.00022$     \\      
 $0.9650$          &           $0.02064\pm0.00021$       &    $0.02071\pm0.00022$     \\      
 $0.9680$          &           $0.02058\pm0.00021$       &    $0.02065\pm0.00023$     \\      
\end{tabular} 
\end{ruledtabular}
\caption{Simulation results for the variation of the inflaton field at horizon crossing for the second and third order of slow roll approximation.
Here the tensor-to-scalar ratio is fixed to $r=0.005$ while the running is fixed to be $\alpha_s=-0.0025$.
In this case, the variation of the field in which $n_s$ appears, become smaller as $n_s$ increases. This is due to the term $1-n_s$. }
\label{tab: table2}
\end{table}

\begin{table}
\begin{ruledtabular}
\begin{tabular}{ccccc}

 $\mu_{\alpha_s}$   &          $\Delta\phi_{h.c}(\Pi_2)$      \\ \hline     
 $-0.0005$          &          $0.02067\pm0.00022$     \\      
 $-0.0015$          &          $0.02069\pm0.00022$      \\      
 $-0.0025$          &          $0.02072\pm0.00022$      \\  
\end{tabular} 
\end{ruledtabular}
\caption{Simulation results for the variation of the inflaton field at horizon crossing for the three different values of the running $\alpha_s$.
In this table we report the results for the higher order expression because is the only one in which $\alpha_s$ appears.
The size of the variation is pushed toward smaller values (in modulus) for very low values of the running.}
\label{tab: table3}
\end{table}

\section{Theoretical implications and Conclusions}

This analysis is performed using an extension of the standard Lyth formula
for the variation of the field about the horizon crossing epoch of quantum modes.
The computation of the expansion has been done likewise the local reconstruction of the effective inflationary potential around $\phi_*$ \cite{36}.
The first step is to introduce higher order terms in the slow roll picture and then, reword them in terms of the cosmological variables.
In particular we derived a third order expansion for $\Delta\phi_{h.c}$.
The Eq.($\ref{eqn: alphadep}$) shows a linear dependence on $\alpha_s$ and it turns out to be decoupled by the other inflationary parameters.
This is because there is not coupling between the third slow roll parameter, $\xi^2$, and the first two slow roll parameters [cfn. Eq.($\ref{eqn: third}$)].
It is straightforward to imagine that a hypothetical fourth order term will introduce a coupling between the running $\alpha_s$ and $r$ and $n_s$ and in addition
it will introduce the higher order variables $d\alpha_s/d\ln k$.
In this work, we performed Monte Carlo simulations assuming possible results of a single foreseen cosmological experiment.
Morever, several new experiments have been proposed (see Sec. I for references).
Therefore, we could combine the related results to get more accurate estimation on 
the main cosmological parameters ($r,n_s,\alpha_s$), on some fundamental reheating variables (like the number of $e$-foldings during reheating stage $N_{reh}$ or the final reheating temperature, $T_{reh}$), and finally on $\Delta\phi_{h.c}$.\\
Mathematically, constraining the variation of the scalar field at horizon crossing is particularly important because provides a lower limit
for the total excursion of the field.
At the same time, we can observe the contribution of the observable quantities to $\Delta\phi$. 
Constraining little variations of $\Delta\phi_{h.c}$ in terms of $M_p$ (or $m_p$), does not imply a sub-Planckian total variation of field.
This is because the last $e$-folds provides a large contribution on the total variation as reported in \cite{3}.
The size of the minimum variation of the scalar field can be related to a hypothetical eternal inflation.
In the simplest scenario, following W.H.Kinney and K.Freese \cite{35}, one can argue that a constant negative running $\alpha_s$ can prevent an eternal inflationary phase to occur.
Since, at third order in the slow roll expansion, the variation of the field around the horizon crossing gets dependence in $\alpha_s$
we find that is possible to split $\Delta\phi_{H.c.}$ in two classes:
one associated with the occurrence of the eternal inflation phase and one not.
However, we should underline a couple of important questions about these results.
First of all we do not have any guarantee that the function Eq.($\ref{eqn: f}$) is well described only by constant $\alpha_s$ (and $n_s$).
In other words, we cannot be sure that a dramatic change (or not) of the order of magnitude about $P(s)$ is only related to the single running $\alpha_s$ parameter.
In principle, the higher order terms may play an important role (see \cite{35}) and so we should generalize the procedure.
On the other side, we cannot measure the running as well as any other cosmological variables outside our horizon, so it is also plausible that
there is no observation can tell us if eternal inflation takes place or not.
From this point of view, it is important to stress that a discussion about the relation between $\Delta\phi_{H.c.}$ and eternal inflation is still speculative.
Surely, one can think that eternal inflation can be occur inside our horizon scale.
In that case we should have observable consequences such as a natural and strong production of black holes.
In our analysis we adopted a single field slow roll version of inflation as the paradigm of the early universe.
This scenario could be confirmed and strengthened in the close future.
Nevertheless there is still room (although little) for non-trivial signature of inflation.
In this respect, the standard definition of Lyth Bound for the minimum variation of the scalar field is not more consistent
as well as any kind of its slow roll generalizations.
Then we need a new definition of $\Delta\phi$ during the horizon crossing of quantum fluctuations.
This problem has been approached for example by Baumann and Green in \cite{37}, in the context of general single field model ($P(X)$-model, \cite{38})
using the Goldstone picture of effective field theory of inflation (see \cite{39}).

\begin{acknowledgments}
This work was supported in part by the STaI ``Uncovering Excellence" grant of the University of Rome
``Tor Vergata", CUP E82L15000300005.
I would like to thanks Paolo Cabella for several discussions and suggestions
during the preparation of the paper.
Many thanks to Gianfranco Pradisi for the support and corrections and the crucial discussion about the general framework of effective field theories
Sabino Matarrese and Nicola Bartolo and for the many advices regarding the early version of the manuscript and Nicola Vittorio for the support.
\end{acknowledgments}




\nocite{*}

\bibliography{apssamp}

\end{document}